\documentstyle[aps,epsfig,float,twocolumn,prl]{revtex}
\newcommand{\be}{\begin{equation}}
\newcommand{\ee}{\end{equation}}
\newcommand{\vi}{\vec i}
\newcommand{\vj}{\vec j}
\begin{document}
\twocolumn[\hsize\textwidth\columnwidth\hsize\csname @twocolumnfalse\endcsname
\draft
\title{Effective Gauge Theories of Spin Systems}
\author{M. B. Hastings}
\address{Center for Nonlinear Studies, MS B258, Los Alamos
National Laboratory, Los Alamos, NM, 87545, hastings@lanl.gov}
\date{October 18, 2000}
\maketitle
\begin{abstract}
A large variety of microscopic gauge theories can be written for
antiferromagnetic spin systems, including $U(1), SU(2)$, and $Z_N$.
I consider the question of the appropriate {\it effective} gauge
theory for such systems.  I show that while an $SU(N)$ anti-ferromagnet can
be written {\it microscopically} as a $Z_N$ gauge theory,
for unfrustrated systems, with a two-sublattice structure, 
there is always an effective $U(1)$ gauge field.  The
dispersion relation for the gauge field is shown to depend on the
presence or absence of charge-conjugation symmetry.
Frustrated systems can break the gauge group to a
discrete group, but this appears to always involve introducing a
gap for the spinons.  
\end{abstract}
]
\section{Introduction}
Spin systems naturally give rise to dynamics on a constrained Hilbert space.
The spin-$1/2$ Heisenberg antiferromagnet arises from a Hubbard model of
fermions, constrained to single occupancy per site by strong on-site
repulsion, while a dimer model leads to a space with each site participating 
in one dimer.  In consequence, it is not surprising that a gauge theory should 
be the correct description.  A variety of effective gauge theories have been
suggested, including the $U(1)$ gauge theory\cite{afm}, the $SU(2)$ gauge
theory\cite{xgw}, and the $Z_2$ gauge theory\cite{xgw,sachdev,senthil}.

It has been shown that the $Z_2$ gauge symmetry is correct for
short-range RVB states on a triangular lattice\cite{sondhi}, and that
there may be a deconfined phase.  A similar argument has been made by 
considering dimer states on the square lattice with diagonal
bonds\cite{sachdev2}.  Finally, a $Z_2$ symmetry has been shown for
the $SU(2)$ model on the square lattice with diagonal bonds\cite{xgw}.
For unfrustrated systems with holes, a 
$Z_2$ gauge theory has again been proposed\cite{senthil}, though it
has been criticized on the grounds of neglecting a $U(1)$ 
symmetry\cite{baskaran}.  Emphasizing the different possible gauge theories
for the same model, while the Rokhsar-Kivelson (RK) 
model on a square lattice can
be described by a $U(1)$ gauge theory\cite{fk,rs1} and has gapless excitations
suggestive of a $U(1)$ theory, there is an exact microscopic
mapping to a $Z_2$ gauge theory\cite{sondhi1}.
The different behavior of the excitations on square and triangular lattices
suggests that discrete gauge symmetries require frustration.

To better understand the correct gauge theory description,
I consider an $SU(N)$ antiferromagnet.  I assume that the ground state
is known, and consider possible excitations above the ground state
in a single-mode approximation.  I show that given an unfrustrated system with a
two-sublattice structure there is always an effective $U(1)$ symmetry,
with the dispersion relation for the excitations depending on the
presence or absence of charge-conjugation symmetry.  In contrast, I show
that {\it microscopically} a $Z_N$ gauge theory suffices, but I point out
the difficulty in the microscopic derivation of the gauge theory.
However, I show that
for frustrated systems one may generate an effective $Z_2$ (or other discrete
group) gauge theory.
\section{Two-Sublattice $U(1)$ Gauge Symmetry in Dimer States}
The Affleck-Marston (AM) large $N$ theory\cite{afm} and the RK
model both have gapless excitations, with differing dispersion relations.
I will show that the AM theory can also be written as a model
of dimers, and that an appropriate generalization of the RK single-mode
operator acting on the AM ground state produces exactly the AM gauge
excitations.  The different dispersion relation is related to the
presence or absence of charge-conjugation symmetry.

Consider a square-lattice, or other two-sublattice, $SU(N)$ antiferromagnet. 
Label the sublattices A and B.  Use the fermionic representation of spins
following Affleck and Marston\cite{afm}.  Consider a Hamiltonian
\be
\label{hamil}
H=-\sum\limits_{\vi,\vj} \Bigl (
\frac{J}{N} |{\psi^{\mu}_{\vi}}^{\dagger} \psi_{\vj}^{\mu}|^2 -
\frac{\tilde J}{N^3} |{\psi_{\vi}^{\mu}}^{\dagger} \psi_{\vj}^{\mu}|^4
\Bigr),
\ee
where the Hamiltonian acts on nearest neighbor lattice sites $\vi,\vj$.
Let $N-m$ fermions be placed on each site in sublattice A and $m$
fermions be placed on each site in sublattice B.
Two possibilities are of interest, depending on $m$.
If $m=N/2$, charge conjugation symmetry is
present and, depending on the ratio $\tilde J/J$,
the large $N$ mean-field solution of Hamiltonian (\ref{hamil}) leads to
either a state with gapless spinons in the $\pi$-flux phase\cite{afm}, or
to a state with spin-Peierls order and gapped excitations.
If $m\neq N/2$, the spinon spectrum is always gapped; for
$m<<N$, there is an effective dimer model\cite{affleck}.

For either possibility, we may make a particle-hole transformation on
the sites in the A sublattice (after transformation, we have an
$Sp(N)$ system\cite{vojta}).
The Hamiltonian becomes
\be
\label{conj}
H=-\sum\limits_{\vi,\vj} \Bigl (
\frac{J}{N} |\psi_{\vi}^{\mu} \psi_{\vj}^{\mu}|^2 -
\frac{\tilde J}{N^3} |\psi_{\vi}^{\mu} \psi_{\vj}^{\mu}|^4
\Bigr).
\ee

Consider a dimer state
\be
\prod\limits_n ({\vi}_n,{\vj}_n)\equiv
(\prod\limits_n (\psi_{\vi_n}^{\mu} \psi_{\vj_n}^{\mu})^{\dagger})|0 \rangle,
\ee
where $|0 \rangle$ is the no-particle vacuum state.
All the sites $\vi$ are on the A sublattice and all the sites
$\vj$ are on the B sublattice.  Each site participates in $m$ dimers.
Hamiltonian (\ref{conj}) acts on a dimer state to produce another
dimer state.  Further, the Hamiltonian preserves the sublattice structure in the
(non-orthogonal) dimer basis.

Rokhsar and Kivelson\cite{rk} noticed the existence of gapless ``resonon"
modes in a dimer model at the RK point where the wavefunction is
an equal amplitude superposition of dimer states.  
These modes may be obtained by
introducing a gauge field $\vec A(\vec x)$, where $\vec x$ is a point on the
dual lattice.  The ground state wavefunction, $\Psi_0$, is a superposition of
different dimer states $\Psi_a$ with amplitudes $A_a$.  Take each dimer state,
\be
\label{pad}
\Psi_a= \prod\limits_{n} (\vi_{n,a},\vj_{n,a}),
\ee
and change the phase of every dimer in that state as
\be
(\vi,\vj) \rightarrow e^{i \vec A(\frac{\vi+\vj}{2}) \cdot (\vi-\vj)} (\vi,\vj).
\ee
In the limit that $A$ is small, the superposition of these dimer
states, orthogonalized with respect to $\Psi_0$, is
\be
\label{dp}
\sum\limits_a A_a \sum\limits_{n'} 
\vec A(\frac{{\vi}_{n',a}+{\vj}_{n',a}}{2}) \cdot (\vi_{n',a}-\vj_{n',a})
\prod\limits_{n} ({\vi}_{n,a},{\vj}_{n,a}).
\ee
Choosing $\vec A(\vec x)=\hat A e^{i \vec k \cdot \vec x}$, with
$\hat A\perp \vec k$, this
is an excitation above the ground state with momentum $k$.  Rokhsar and Kivelson
showed that this mode has dispersion $E\propto k^2$, under
the assumptions that the ground state $\Psi_0$ was an equal amplitude
superposition of dimer states and that the
overlap between dimer states can be ignored (valid in the large $N$
limit).

Similarly, Affleck and Marston found gapless $U(1)$
gauge fields about the large $N$ flux-phase solution with $m=N/2$.
In the flux-phase, the Hamiltonian (\ref{hamil}) is
decoupled by a field $t_{ij}$; the mean-field solution is described by
the hopping Hamiltonian
\be
\label{hamf}
H=\sum\limits_{\vi,\vj}
t^{\rm flux}_{\vi,\vj} {\psi_{\vi}^{\mu}}^{\dagger} \psi_{\vj}^{\mu}+ {\rm h.c.}
\ee
with a mean-field solution $t^{\rm flux}$.  After
the particle-hole transformation, the ground state of Hamiltonian
(\ref{hamf}) is described by a collection of Cooper pairs with
particles in the pair on opposite sublattices.  Therefore, the
solution of Hamiltonian (\ref{hamf}) can be described
as a superposition of dimer states $\Psi_a$ given by
equation (\ref{pad}), with amplitudes $A_a$; due to the
mean-field nature of the solution, the number of dimers that each
lattice site participates in is not fixed.

For $m\neq N/2$, the mean-field Hamiltonian (\ref{hamf}) has an additional
chemical potential term, alternating on the two sublattices.
After the particle-hole transformation, this state, too,
can be described by a projection of dimer states.  In the limit of
large chemical potential, the AM state evolves into the ground state of the
RK point.

The fluctuations in the amplitude of $t$ are gapped, but there are
gapless phase fluctuations, which give rise to a $U(1)$ gauge field.
It may be shown that the solution of the mean-field
Hamiltonian in the presence of a transverse gauge field\cite{itz}
reduces to, for weak fields, a superposition of states given by equation 
(\ref{dp}).  Intuitively this may be understood as follows: a fluctuation in 
$t$ which is pure gauge, so that
\be
t_{\vi,\vj}=t_{\vi,\vj}^{\rm flux} e^{i (\theta_{\vi}-\theta_{\vj})},
\ee
multiplies the Green's function by
$e^{i (\theta_{\vi}-\theta_{\vj})}$ and hence multiplies each
dimer by $e^{i (\theta_{\vi}-\theta_{\vj})}$.  Taking 
$\theta_{\vi}=\vec A \cdot \vi$, and then letting $A$ become space-dependent
yields equation (\ref{dp}).  
Thus, not only does the AM ground state evolve into the RK ground state
as $m$ varies, but the gauge excitations also are connected.

To elucidate why the AM gauge field has 
$E \propto |k|$ while the RK gauge field has
$E \propto k^2$, consider excitations in single-mode
approximation about an arbitrary dimer
state, assumed to be the ground state of some unspecified Hamiltonian
that preserves the sublattice structure of dimer states.
Take a gauge field excitation at wavevector $k$, while 
the dimers have range $l$.
The dispersion of the gauge field will depend on the presence
or absence of charge conjugation symmetry.

For a given dimer state, $\Psi_a$, define the ``current" $\vec J(\vi)$, for
$\vi$ on sublattice $A$, to be equal to the sum over sites $\vj$ on sublattice
$B$, such that $\vi$ and $\vj$ are connected by a dimer, of $\vi-\vj$.
Then, the state (\ref{dp}) can be obtained from the ground state by
acting with operator
\be
\label{dop}
O(\vec k,\hat A)=
\sum\limits_{\vi} \hat A \cdot \vec J(\vi) e^{i \vec k \cdot \vec x}.
\ee
The energy of the excited state is equal to $f(O)/s(O)$, where
the oscillator strength $f(O)=\frac{1}{2}\langle \Psi_0 |
[O^{\dagger},[H,O]]|\Psi_0 \rangle$, and the structure factor
$s(0)=\langle \Psi_0| O^{\dagger} O |\Psi_0 \rangle$.

If we sum the current $\vec J$,
over an area $X$ with length scale much longer than $l$, we obtain
\be
\label{cons}
\sum\limits_{\vi} \vec J(\vi)
= \sum\limits_{\vj} N(\vj) \vj -\sum\limits_{\vi} 
N(\vi) \vi,
\ee
where the sums extend over $\vi,\vj$ in $X$ and $N(\vi),N(\vj)$ is
defined to be the number of dimers which connect $\vi$ or $\vj$ to
a site outside of $X$.  Eq. (\ref{cons}) implies a conservation
law for the current on length scales greater than $l$, as 
nonvanishing contributions to the total current arise only for
$\vi,\vj$ near the boundary of $X$.

First we consider $m<<N$, 
so that the overlap between dimer states may be ignored.
Overlap produces ferromagnetic correlations
between spins on the same sublattice, absent in the infinite $N$ limit.
The overlap between dimer states leads to a loop gas\cite{lg}; for
large $l$ and small $N$ there is a phase with arbitrarily long loops.
The most likely result of this phase transition is 
long range N\'eel order.
Restricting our attention to systems without long-range
order permits us to ignore this possibility.

Suppose $k^{-1}>>l$.  The oscillator strength $f(O)$ must vary as $(\vec k\times
\hat A)^2$: the vanishing of $f(O)$ for longitudinal excitations is
due to current conservation, while the $k^2$ variation is required by
analyticity and $f(O)=0$ at $k=0$.

The longitudinal structure factor vanishes for small $k$.
The transverse structure factor vanishes exactly for certain translational
symmetry-breaking states, such as a columnar state in a dimer model,
as well as for C-breaking states\cite{cbreak} in which,
after the particle-hole transformation,
each lattice site participates in $m/4$ dimers with each of its
four neighbors.  Even in the event of weak fluctuations about
these states, if the long-range dimer order is preserved the structure
factor vanishes as $k^2$, leading to a gap to gauge modes.
Conversely, if the transverse structure factor vanishes
at $k=0$, on length scales much greater than $l$
there is no net current.
If two holon excitations are created in the system on
opposite sublattices (these consist of lattice sites with $m-1$ dimers),
current conservation requires that there be a net current of 1 along
a string connecting the two sites, which implies that along the
string the system is in an excited state, so that the
energy of the state is proportional to the spinon separation.
Thus, we argue that the vanishing of the transverse structure factor implies
confinement of holons.  However, a nonvanishing structure factor implies
gapless gauge excitations with $E\propto |k|^2$.

Now we consider $m\approx N/2$; since overlap in the dimer basis
becomes important, we use the fermion basis.
This gives the dispersion for the AM gauge field,
taking the ground state to be fermions with eq. (\ref{hamf}), and
excitation given by eq. (\ref{dp}).  For $m=N/2$ we will obtain
the same $E \propto k$ dispersion relation as was found previously
by integrating over the fermions:
the oscillator strength is proportional to $(\vec k\times \hat A)^2$,
but the structure factor will be proportional to $|k|$.

The operator that creates state (\ref{dp}) from the ground state is
\begin{eqnarray}
O(k)=
\vec A(\frac{{\vi}+{\vj}}{2}) \cdot (\vi-\vj)
\times\nonumber \\
a^{\dagger}(\vi) a^{\dagger}(\vj) a(\vi') a(\vj')
\overline \psi(\vi,\vj)
\psi(\vi',\vj').
\end{eqnarray}
Here $\psi$ is the Cooper pair wavefunction.  
Undoing the particle-hole transformation so that there is a Fermi sea,
the destruction of the Cooper pair creates
a particle-hole excitation at opposite momenta: $k_1,-k_1$.  The
creation of the Cooper pair creates a particle-hole excitation at
momenta $k_2+k/2,-k_2+k/2$.  
Both $k_1,k_2$ lie within the Fermi sea.  Therefore,
$k_1=\pm k_2+k/2$, with $-k_1$ within the Fermi sea and
$\mp k_2+k/2$ outside the Fermi sea, so that $O(k)$ creates one particle
and one hole.  The phase space volume for this is proportional to $|k|$,
giving the desired result for the structure factor.
To reconcile this result
with the nonvanishing structure factor found above
in the dimer basis, one must include overlap; the structure factor of $|k|$ 
found in the fermion basis for $m=N/2$ is a result of cancellations due to 
overlap.

For $m\neq N/2$, there are two separate bands, and,
using the fermion basis, the structure factor is 
constant for small $k$, leading
to $E \propto k^2$.  
Equivalently, the cancellations due to overlap in the dimer basis
are no longer exact for $m\neq N/2$ and
the structure factor becomes constant.
For $m\neq N/2$, after integrating over fermions
the effective action for the
AM gauge field has a Maxwell term, so that the quadratic dispersion relation is
surprising; however, the gauge field is coupled
to an alternating charge density on each site, producing a
term in the action proportional $A_t(\vi)-A_t(\vj)$, the
difference in the time component of the gauge field on the two sublattices.
Such a term would otherwise be forbidden by charge-conjugation symmetry.
As a result, there is
a flux in the gauge field from sites on the A sublattice to the B sublattice.
The dynamics in this space of states, in which there is a given amount
of flux leaving each site on the $A$ sublattice going to a site on
the $B$ sublattice, is closely related to the dynamics of a dimer model
and gives the $k^2$ dispersion.
\section{Microscopic Gauge Theories}
Despite the $U(1)$ gauge symmetry of the effective theory of an unfrustrated
anti-ferromagnet, the microscopic
field theory may be written as a $Z_N$ gauge theory.
Formally, a system with Hamiltonian $H$ given
by equation (\ref{hamil}) can be written as a functional integral
\be
Z=\int [{\rm d}\overline\psi^{\mu}][{\rm d}\psi^{\mu}][{\rm d}t_{ij}]
e^{-S[\overline \psi,\psi,t_{ij}]},
\ee
with action
\be
S=\int 
\sum\limits_{ij} 
\overline \psi^{\mu}_i(\partial_t \delta_{ij}+t_{ij} )\psi^{\mu}_j 
+ f(|t_{ij}|)
{\rm d}t.
\ee
for an appropriate function $f$.  The decoupling field, $t$, obeys
$t_{ij}=\overline t_{ji}$.
However, in contrast to the large $N$ AM gauge theory\cite{afm},
for finite $N$ the above decomposition is still 
valid if the variables $t_{ij}$ take
only the values $t_{ij}=|t_{ij}| e^{i \theta_{ij}}$, with
$\theta$ an integer multiple of $2 \pi/N$.

Then, by integrating out the fermions, a plaquette action is induced
for $t$, yielding a $Z_N$ gauge theory.  However, as argued
above, there is always a $U(1)$ symmetry present for such an antiferromagnet
on a bipartite lattice.  In consequence, the microscopic derivation of
the gauge theory should not be trusted.

The field $t_{ij}$ is conjugate to $\overline \psi^{\mu}_i \psi^{\mu}_j$.
Within an $SU(N)$ theory, one finds that $(t_{ij})^{N}$ vanishes
within the $m$-particle-per-site Hilbert space for any $m<N$.  
This is the formal device that lets one replace $t$ with a $Z_N$ gauge field.  
However, plaquette operators for $t$ are generated under an RG.  
There are two possible forms of these.  One is $t_{ij} t_{jk} t_{kl} t_{li}$ 
around a plaquette.  For $SU(2)$, this operator is self-cube\cite{jbm}, which 
suggests a $Z_2$ gauge theory.  However, another, more symmetric, plaquette term
is $(S_i^{\mu\nu} S_j^{\nu\rho} S_k^{\rho\sigma} S_l^{\sigma\mu})$,
where the spin operator $S=
\overline \psi^{\mu}_i \psi^{\nu}_i-\frac{1}{2}\delta^{\mu\nu}$.
This operator cyclically permutes the spins on sites $i,j,k,l$.  The
fourth power of this operator is equal to unity, which suggests that
the gauge symmetry is at least $Z_4$; by continuing to larger traces of
spin operators, the full $U(1)$ symmetry is restored.
\section{Frustrated Antiferromagnets}
In contrast to this result for unfrustrated antiferromagnets,
a frustrated antiferromagnet may have a discrete gauge symmetry.
Due to the lack of a two-sublattice structure, the type of
$U(1)$ symmetry considered above is not possible.

On the two-sublattice systems, one may take conjugate representations of
$SU(N)$ on alternate sublattices.
For a frustrated system, this is
not possible, and one must either use the group $Sp(N)$\cite{sachdev},
or use the self-conjugate $SU(N)$ system with $m=N/2$.
The two-sublattice $U(1)$ symmetry we consider above is a $U(1)$ symmetry
between conjugate representations of $SU(N)$; for a frustrated $SU(2)$
system, in addition to forming singlet operators 
${\psi_{\vec i}^{\mu}}^{\dagger} \psi_{\vec j}^{\mu}$ between sites 
$\vec i,\vec j$ on opposite sublattices, one may also form singlets 
$\epsilon^{\mu\nu} {\psi_{\vec i}^{\mu}}^{\dagger}
{\psi_{\vec j}^{\nu}}^{\dagger}$ 
between sites $\vec i,\vec j$ on the same sublattice, destroying the
$U(1)$ symmetry.  In general, the symmetry is broken to $Z_N$.

Examples of effective $Z_2$ theories include
the dimer model on the triangular lattice\cite{sondhi} and
the frustrated square lattice $SU(2)$ theory\cite{xgw}.
These theories lead to gapped,
deconfined spinons.  No
frustrated system with gapless spinons has been found.  From the
point of view of the $SU(2)$ mean-field theory, this is to be expected.
For the unfrustrated square-lattice, the $\pi$-flux phase enlarges the
unit cell of the lattice to two sites, so there
is an integer number of each flavor of spinon per unit cell.  One would
expect such a system to be a band insulator; however, the two-sublattice 
structure leads to a zero-energy state and keeps the spinons gapless.  As soon
as this structure is lost, the spinons are expected to
become gapped.
\section{Conclusion}
The effective gauge theories that describe spin systems have been
considered from excitations above the ground state.  
I have argued that, despite the possibility
of a microscopic $Z_N$ gauge theory, there is always an effective $U(1)$
symmetry for unfrustrated spin systems.  While for frustrated spin
systems an effective $Z_2$, or other discrete gauge symmetry, is possible, this
appears to coincide with a gap to spinon excitations.  

The discussion above, however, is specific to spin systems.  It is
possible for systems with holes to break the symmetry to $Z_2$
as suggested by Senthil and Fisher\cite{senthil}.

The effect of the $U(1)$ gauge excitations is unclear.  While Marston\cite{inst}
has argued that instantons in the $U(1)$ gauge theory of fermionic $SU(N)$
antiferromagnets do not necessarily lead to 
confinement, another possibility is that gauge theories of two
dimensional antiferromagnets with gapless, deconfined spinons occur only as
theories of a critical point, such as the RK point.
In this case, while some examples
may be found of gapless spinons in two-dimensions, there will always
be relevant perturbations that will lead to some form of ordering and
to spinon confinement.
\section{Acknowledgements}
I would like to thank Shivaji Sondhi, Roderich Moessner,
and Brad Marston for many useful discussions.
I acknowledge support from DOE W-7405-ENG-36.

\end{document}